\documentclass[manuscript]{aastex}
\usepackage{lscape}
\usepackage[3D]{movie15}
\usepackage{hyperref}

\shorttitle{H$\alpha$ observations of a partial CME}
\shortauthors{D. Christian et~al.}

\begin{document}

\title{H$\alpha$ and EUV observations of a partial CME}

\author{Damian J. Christian}
\affil{Department of Physics and Astronomy,  
California State University Northridge, 18111 Nordhoff Street, Northridge, CA 91330, USA; daman.christian@csun.edu}

\author{David B. Jess}
\affil{Astrophysics Research Centre, School of Mathematics and Physics, Queen's University Belfast, Belfast, BT7 1NN, Northern Ireland, U.K.}

\author{Patrick Antolin}
\affil{National Astronomical Observatory of Japan, 2-21-1 Osawa, Mitaka, Tokyo 181-8588, Japan
}

\author{Mihalis Mathioudakis} 
\affil{Astrophysics Research Centre, School of Mathematics and Physics, Queen's University Belfast, Belfast, BT7 1NN, Northern Ireland, U.K.}



\begin{abstract}

We have obtained H$\alpha$ high spatial and time resolution observations of the upper solar chromosphere and supplemented these with multi-wavelength observations from the Solar Dynamic Observatory (SDO) and the {\it Hinode} ExtremeUltraviolet Imaging Spectrometer (EIS). The H$\alpha$ observations were conducted on 11 February 2012 with the Hydrogen-Alpha Rapid Dynamics Camera (HARDcam) instrument at the National Solar Observatory's Dunn Solar Telescope.  Our H$\alpha$ 
observations found large downflows of chromospheric material returning from coronal heights following a failed prominence eruption.
We have detected several large condensations (``blobs'') returning to the solar surface at velocities of $\approx$200 km s$^{-1}$ in both H$\alpha$ and several SDO AIA band passes. 
The average derived size of these ``blobs'' in H$\alpha$  is 500 by 3000 km$^2$ in the directions perpendicular and parallel to the direction of travel, respectively. 
A comparison of our ``blob" widths to those found from coronal rain,  indicate there are additional smaller, unresolved ``blobs" in agreement with previous studies and recent numerical simulations.
Our observed velocities and decelerations of the ``blobs'' in both H$\alpha$ and SDO bands are less than those expected for gravitational free-fall and
imply additional magnetic or gas pressure impeding the flow.
We derived a kinetic energy $\approx$2 orders of magnitude lower for the main eruption  than a typical CME, which may explain its partial nature.

\end{abstract}


\section{Introduction}
\label{intro}


Building on the success of studies with SoHO and TRACE, the Solar Dynamic Observatory (SDO) has uncovered a wealth of new information surrounding the magnetic structuring of the solar atmosphere. These magnetic features, including coronal loops, are found to be especially dynamic, often displaying eruptions related to emerging magnetic flux ropes.
The emergent flux manifests itself as ribbons, loops and strands \citep{MS06}, and
are sometimes accompanied by eruptive events and coronal mass ejections 
(CMEs).
Higher spatial resolution observations of the small-scale structures that arise in the aftermath of CME events, including those related to coronal rain, offer a unique ability to constrain and understand the dynamic processes embedded within the coronal magnetic fields. 

Flares observed with CMEs have been noted as ``eruptive"  or ``full eruptive" events, and flares with no CMEs have been noted as ``confined" or ``failed" \citep{WZ07, K13}.
In a full CME, the plasma and associated magnetic structure is ejected and escapes from the Sun, while in a failed CME the plasma does not escape and falls back \citep{G07, M11}.
More recent observations have found, that regardless of whether the CME is a ``full" or ``failed" event, some material from the filament/prominence is often observed returning back to the solar surface \citep{I12,G13}.
\citet{G07} also define a ``partial" filament eruption with two classes, A and B. The class A partial eruption has an eruption of the entire magnetic structure with a small amount of mass. Only part of the magnetic structure is observed to erupt in Class B, also with a small amount of mass. \citet{M11} reviews several of the mechanisms, other than solar gravity, that can explain failed CMEs. 

Recent observational studies of CMEs have found behavior and activity that can be interpreted by ``tether cutting",  ``magnetic reconnection" or MHD kink instability scenarios. There are even instances where more than one of the scenarios are needed to explain the observations \citep{WTD05}. 
\citet{R10} found support for the``tether-cutting" mechanism using  multi-spaceraft X-ray observations of a C8 class flare plus CME. Recently, \citet{Bi13} 
found rotation and non-radial motion during an eruptive filament.
Many previous studies also found evidence for magnetic reconnection 
\citep{M01, GM04, J07, S12}.
Additionally, \citet{KI13} 
found evidence for blast waves leading to the break-out of the flux rope after magnetic reconnection, while \citet{T07} 
also observed magnetic reconnection inside an emerging flux rope for a bright coronal downflow after a CME. There is also a heated debate on whether or not the magnetic 
flux rope exists before the onset of the CME or is formed during the process \citep{P13}. 

Although failed CMEs have been poorly studied,  \citet{J03}
found evidence for magnetic reconnection in a failed eruption, and \citet{S12}
used multi-angle observations of several failed eruptions to ascribe failed eruptions to several factors, including: stronger magnetic field at low altitudes, low magnetic field gradients of the overlying loops with height, asymmetric magnetic confinement of the overlying fields and the kinetic energy of the erupting filament mass.
Within the class of chromospheric material falling back to the solar surface from coronal heights it is important to distinguish prominence material from coronal rain, which consists of cool plasma condensation rapidly produced in the corona (in a timescale of minutes), falling toward the solar surface along coronal loops 
\citep{K70, L72, S01, dG04,dG05}.
 Coronal rain condensations can have velocities over 100 km/s, with typical velocities of 60-70 km/s  \citep{A10, AVR12a, ARdV12b},
and have been observed to cool very rapidly from 
coronal to chromospheric temperatures \citep{A99}. 
There have been several 
investigations into the agents responsible for  the less than free-fall speed observed in the rain, such as gas pressure \citep{M03, M05, A10, O14},
and ponderomotive force from transverse MHD waves \citep{AV11}.
High temporal and spatial resolution observations of failed CMEs and coronal rain are needed to help distinguish between these different models and scenarios, and promise to help constrain parameters of the solar atmosphere, such as magnetic field strength, gradient, and plasma density. 

In the current paper we present new  H$\alpha$ 
observations  
showing large downflows of chromospheric material returning from coronal heights following a failed prominence eruption.  
These observations were conducted in February 2012 with the 
new Hydrogen-Alpha Rapid Dynamics camera (HARDcam; Jess et al. 2012) instrument and the National Solar Observatory's Dunn Solar Telescope.  We supplemented our optical observations with EUV data obtained with the Solar Dynamic Observatory's (SDO) Atmospheric Imaging Assembly (AIA) and the {\it Hinode} ExtremeUltraviolet Imaging Spectrometer (EIS). Our observations, and analysis are presented in \S\ 2. Our results are given in \S\ 3 and discussed in \S\ 4 with concluding remarks presented in \S\ 5.
 
\section{Observations \& Data Analysis}
\subsection{Observations}
\subsubsection{Ground Based: HARDcam}
The high velocities and relatively small structures in coronal loops require observations with high temporal and spatial resolution and these have only recently been achieved with newer rapid read-out ground-based camera systems, 
and high order adaptive optics for image reconstruction. We obtained
observations of the solar limb at 32N and 85.5E on 2012 February 11 in two sequences at 16:11 and 16:21 UT with  the new H$\alpha$ camera HARDcam \citep{J12} at the Dunn Solar Telescope. HARDcam was run in conjunction with the Rapid Oscillations in the Solar Atmosphere (ROSA,  \citealt{J10}), 
camera system, which was observing photospheric bands not used in the present work.
Our limb observations were conducted as part of our program to observe solar flares (DST proposal number T926), and although we missed the flare near the solar limb, we detected several condensations (``blobs'') returning to the solar surface.

The Hydrogen-Alpha Rapid Dynamics camera (HARDcam, \citealt{J12})  
camera is an electron-multiplying CCD, with a quantum efficiency exceeding 95\% at 6500 \AA\ and is Andor Technology model  iXon X3 DU-887-BV. 
HARDcam has 512$\times$512 pixels and was set-up with a spatial sampling of 0.138 arcsec per pixel, providing a field of view of 71$\arcsec$ x 71$\arcsec$.
HARDcam was used with a 0.25 \AA\ H$\alpha$ core filter and triggered at a constant cadence of 0.05 sec. 

\begin{figure}[t]
\includegraphics[scale=0.66, angle=90]{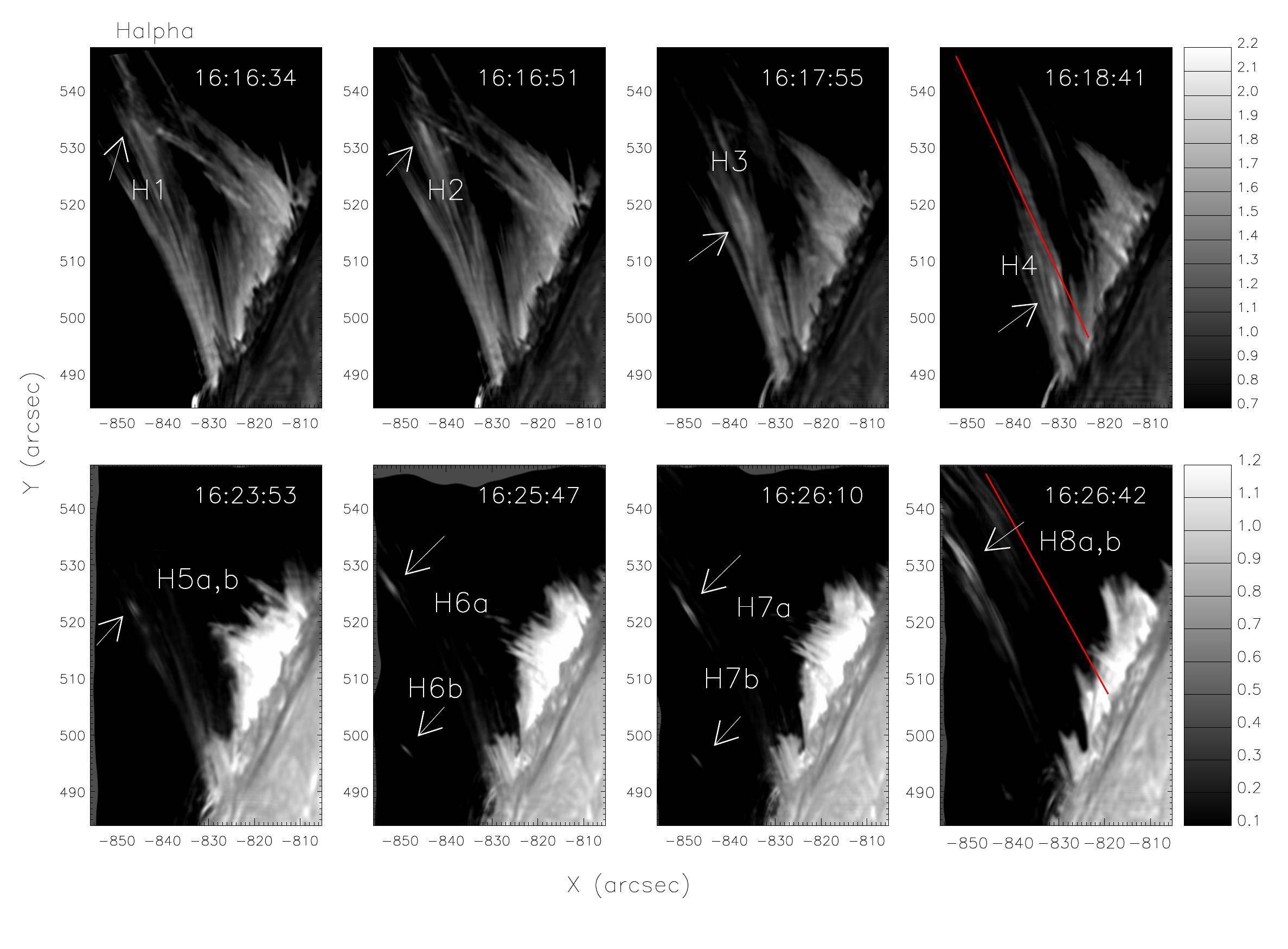}
\medskip
\caption{ HARDcam H$\alpha$  images from 11 Feb 2012 showing blobs of material 
falling back toward the solar surface (see text). Top panel shows selected images for  the first image sequence starting at 16:11:24 UT (Series 1) and the lower panel  shows selected images from the second sequence starting at 16:21:42 UT (Series 2).  
Features of selected blobs can be found in Table 1
The lower right panels for each H$\alpha$ sequence also shows the trace for the X-T cuts used in Figure~\ref{xtcut}.
}
\label{hardcam1}
\end{figure}

All data were obtained using real-time adaptive optics to correct distortions to the wave front and correct the seeing  \citep{R04}.  
Further improvements were applied to the images in processing with speckle reconstruction \citep{W05, WLR05}
and H$\alpha$ images were combined from 35 $\rightarrow$ 1, providing a final image cadence equal to 1.78s.
Images of our 2 H$\alpha$ sequences are shown in  Figure~\ref{hardcam1}.
 
\subsubsection{SDO AIA}
\begin{figure*}
\includemovie[
 controls=true,
text={\includegraphics[width=160mm]{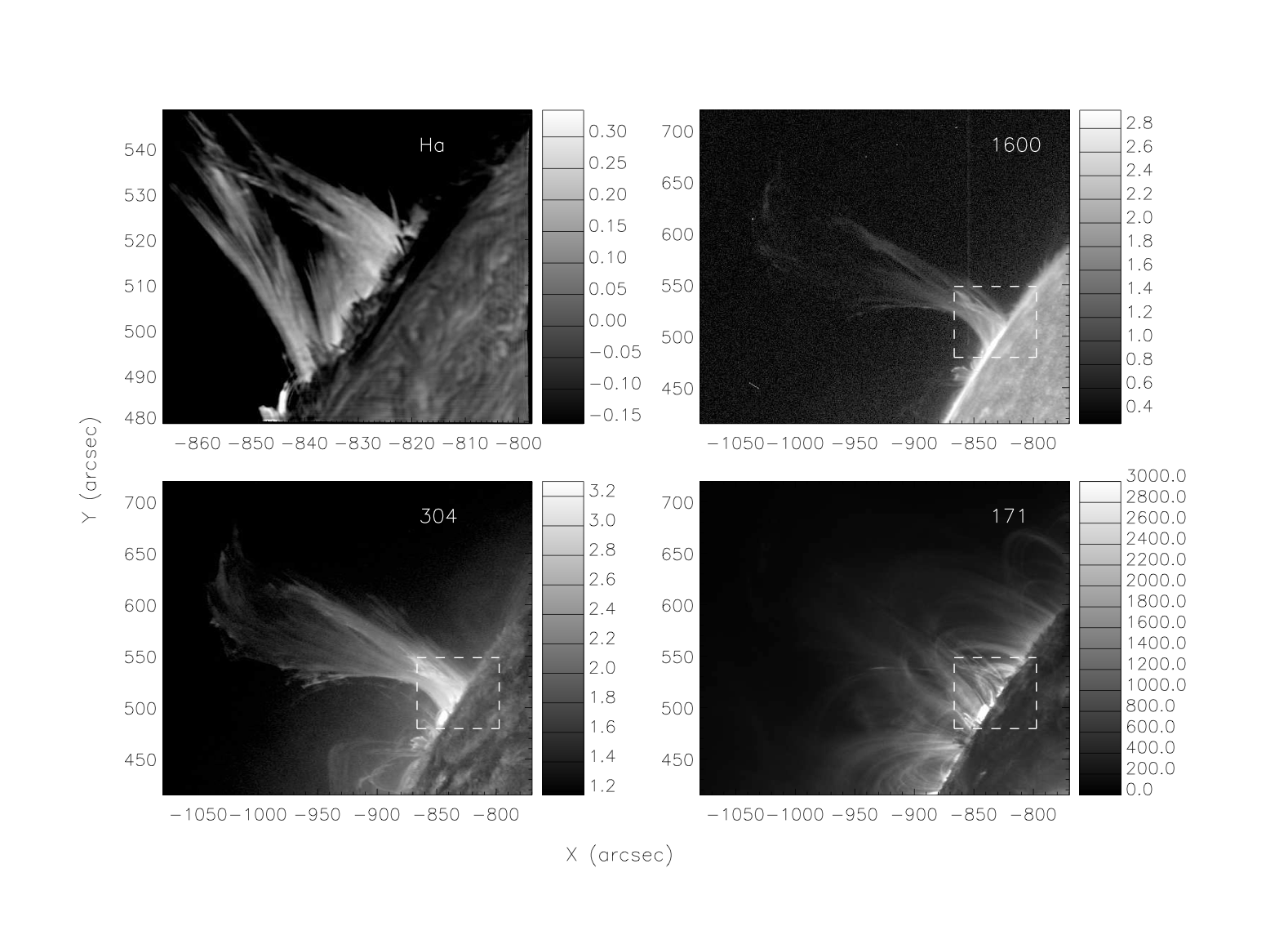}}]
 {\linewidth}{0.5\linewidth}{movie1.mp4}
\caption{HARDcam H$\alpha$ and SDO AIA sample images from Movie 1 in the on-line material. The images are  taken near time 16:16:50 UT.  Shown are the H$\alpha$ HARDcam data in the upper left panel, and SDO AIA  1600, 304, and  171 \AA\ bands in the upper right and lower left and right panels, respectively. All images are displayed using a log scale and the HARDcam field-of-view is indicated in the SDO images as the white box. 
In the movie, the SDO data spans from 15:00 to 17:00 UT and the HARDcam data from $\approx$16:11 to 16:30, and  black frames are shown for the times where there were no HARDCam observations. }
\label{sdohalph4}
\end{figure*}

\begin{figure}
\includegraphics[scale=0.66, angle=90]{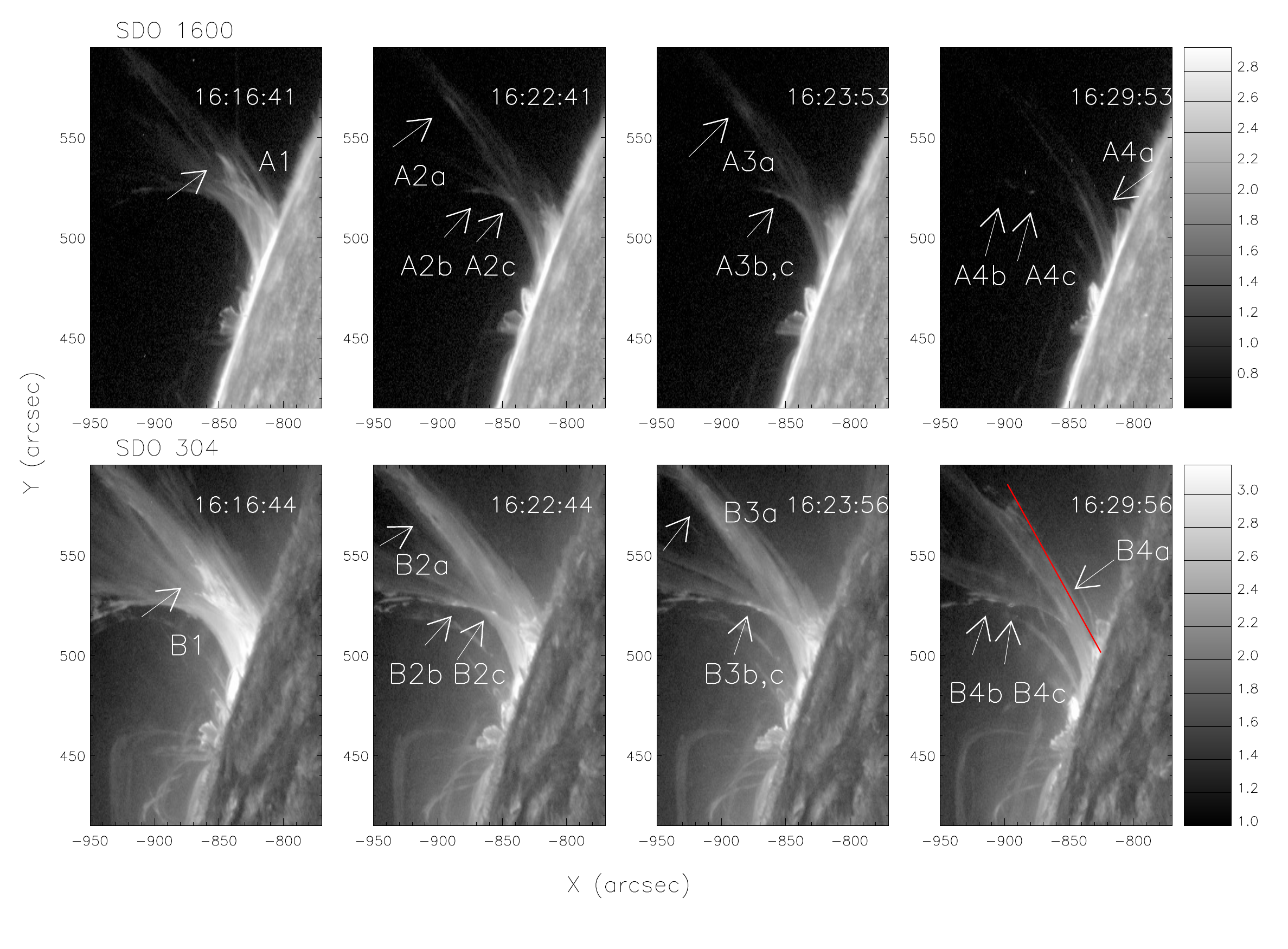}
\caption{SDO AIA sequence of  images from 11 Feb 2012 (near 16:16 UT) showing selected features (``blobs'') of interest.  The 1600 \AA\ and the 304 \AA\ bands are shown in the  {\it  top} and {\it bottom} panels, respectively.  Properties of these blobs are given in Table~1. The lower right panel for the SDO 304 \AA\ image sequence also shows the trace for the X-T cuts used in Figure~\ref{xtcut}.
}
\label{sdo304}
\end{figure}

 We have supplemented our H$\alpha$ observations with EUV
images from the Atmospheric Imaging Assembly (AIA; \citealt{L11}) on-board the Solar Dynamics Observatory (SDO; \citealt{PTC12}). 
 The AIA instrument images the entire solar disk in 10 different channels, incorporating a two-pixel spatial resolution of $1{\,}.{\!\!}{\arcsec}2$ ($\approx$900 km for the AIA's PSF) and a cadence of 12 sec for the EUV channels and 24 sec for the 1600 \AA\ channel. Here, we
selected 5 EUV datasets spanning 15:00 -- 17:00 UT on 11 Feb
2012, consisting of 600 images in each of the 94 , 171,
193, 304 \AA\ channels and 300 images for the 1600 \AA\ channel. 
The SDO observations caught an eruption starting at $\approx$15:55 UT and the subsequent material returning to the solar surface. The main eruption stops its outward expansion and starts to fall back to the solar surface at $\approx$16:11. A smaller eruption is observed to the South-East and starts at $\approx$16:11 and then returns to the solar surface at $\approx$16:17.   GOES detected a C2.7 flare just after 15:55~UT 
and a CME was observed by LASCO C2 at 16:48~UT, but was noted as a ``poor event".  
A sample of images from several SDO bandpasses and our H$\alpha$ data are shown in  (Figure~\ref{sdohalph4}) near the time of the onset of a selected ``blob" in H$\alpha$, and a movie, M1 
is provided in the on-line material for the  H$\alpha$, and SDO AIA 1600, 304 and 171 \AA\ data sets. 
Image sequences for the 304 \AA\ and 1600\AA\ bandpasses are also shown in Figure~\ref{sdo304} with selected features noted and also presented in Table~1. 
 
%
%

\subsubsection{Hinode EIS}
\begin{figure}[t]
\includegraphics[scale=0.66, angle=0]{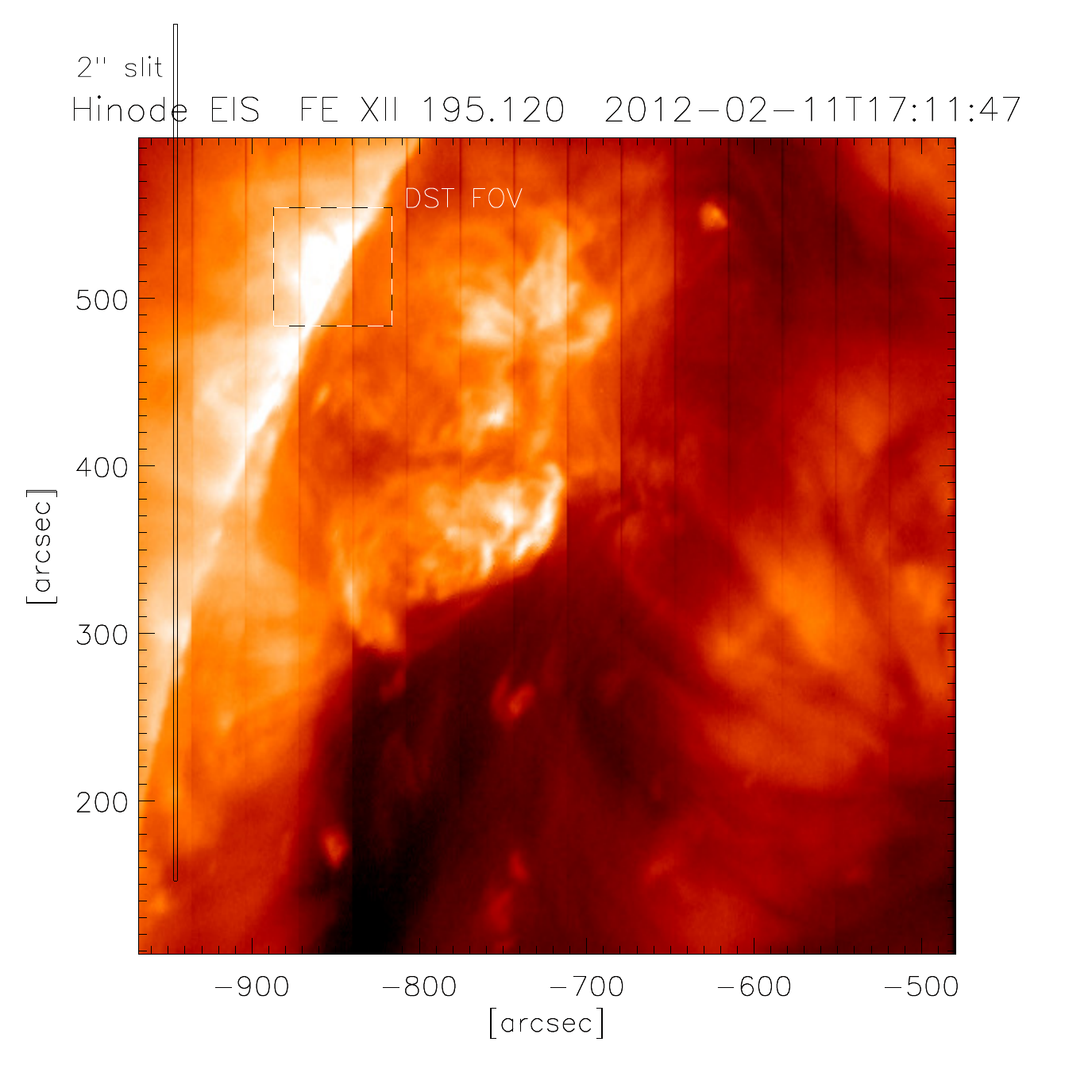}
\caption{{\it Hinode} EIS  Fe XII 195.12 \AA\ image.  The location of the DST field-of-view (white square) as well as that of the EIS 2$\arcsec$ slit is shown.
}
\label{EIS1}
\end{figure}

\begin{figure}
\includegraphics[scale=0.66, angle=0] {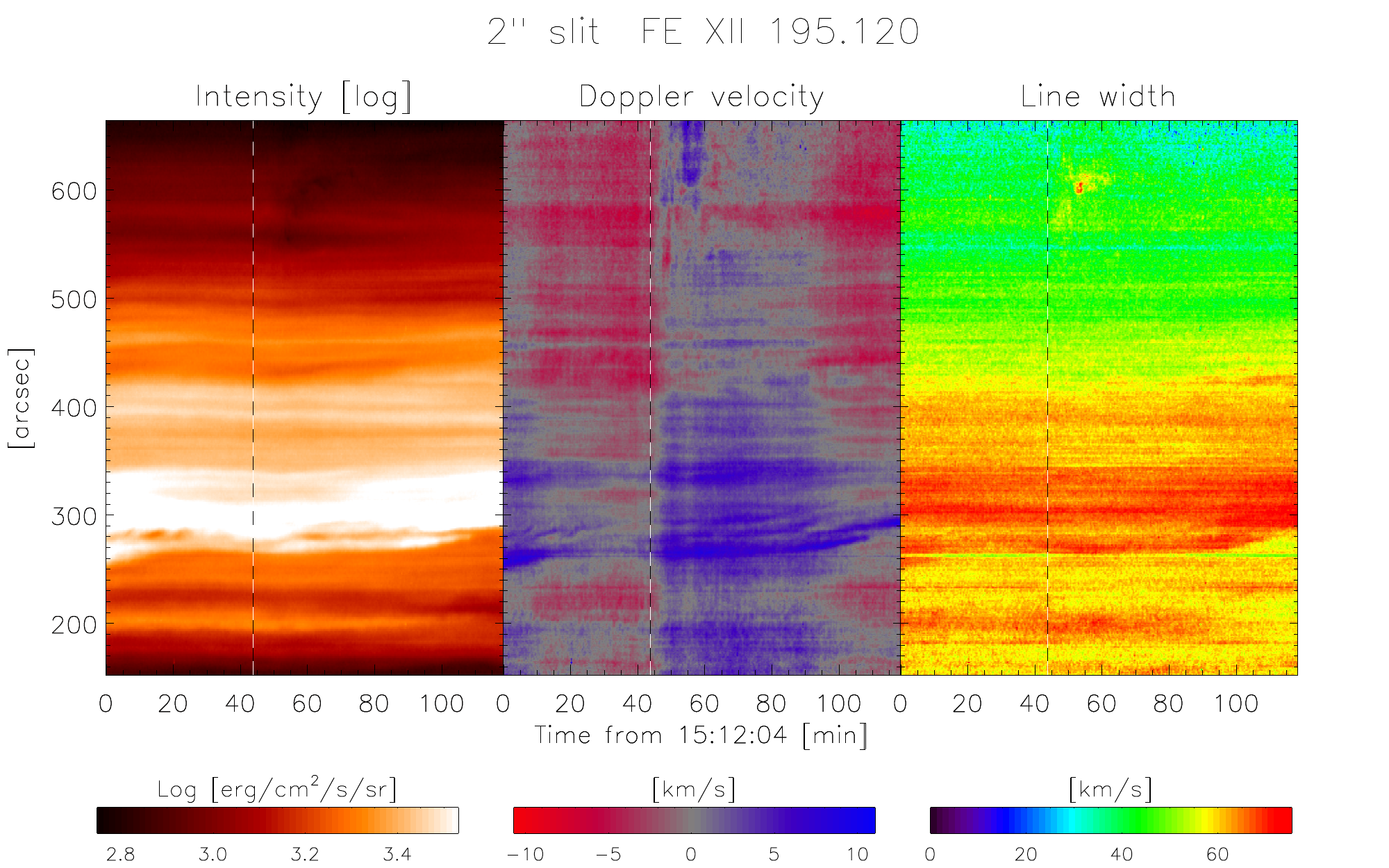}
\caption{{\it Hinode} EIS data (from left-to-right) for: time-distance diagrams for the intensity, Doppler velocity, and line width for the Fe XII 195.12 line, for the time interval of 15:12:04 to 17:10:30 UT.  The zero point of the Doppler velocity is set from the average for a small region of 5 arcsec at the bottom of the slit where no activity was observed, and restricted to 45 min before the eruption. The start of the eruption, as seen in the AIA filters, is marked in 
all panels as a vertical dashed line, and the largest changes after the eruption are seen above the 550$\arcsec$ slit position.
}
\label{EIS2}
\end{figure}

Observations from ExtremeUltraviolet Imaging Spectrometer (EIS; \citealt{C07}) of the Hinode satellite \citep{K07} during the time of our HARDcam observations are also included.  In Figures~\ref{EIS1} and \ref{EIS2} 
we show EIS data at the wavelength of Fe XII 195.12 \AA, formed at a temperature of log T = 6.2. The data were reduced using standard EIS software included in the SolarSoft package \citep{FH98}. The slot image corresponds to a 40$\arcsec$ slot reconstruction at 15 adjacent positions, leading to an entire field of view of 487 $\times$ 487 arcsec$^{2}$, and was taken at 17:11:47 UT on 2012 February 11. 
The slit carried out sit-and-stare observations in Fe XII 186.74 and 195.12, and Fe XVI 262.98, throughout the day in tracking mode with a cadence of 23.8 s. Most of the slit was off-limb throughout the observations, as shown in the context slot figure.
The calculation of the moments for the slit data was carried out using the eis\_auto\_fit
routine, which automatically takes care of the spectrum drift and the slit tilt using the
method described by \citet{K10}. 
\citet{K10} estimate that this method provides an accuracy of $\pm$ 4.4 km/s. On the other hand, the line profiles over the region of interest (off-limb) retain a gaussian shape throughout the observations and therefore we believe single gaussian fits to the data provides good estimates of the line widths regardless of the absolute centroid for Fe XII 195.12 \AA. 
In Figure~\ref{EIS2} 
we show time-distance diagrams for the intensity, Doppler velocity, and line width for the Fe XII 195.12 \AA\ line, for the time interval of 15:12:04 to 17:10:30 UT. The other 2 lines require longer exposure times for proper signal-to-noise and are therefore not included in the present study. 

\begin{figure}
\vspace{0.2in}
\includegraphics[scale=0.50, angle=0]{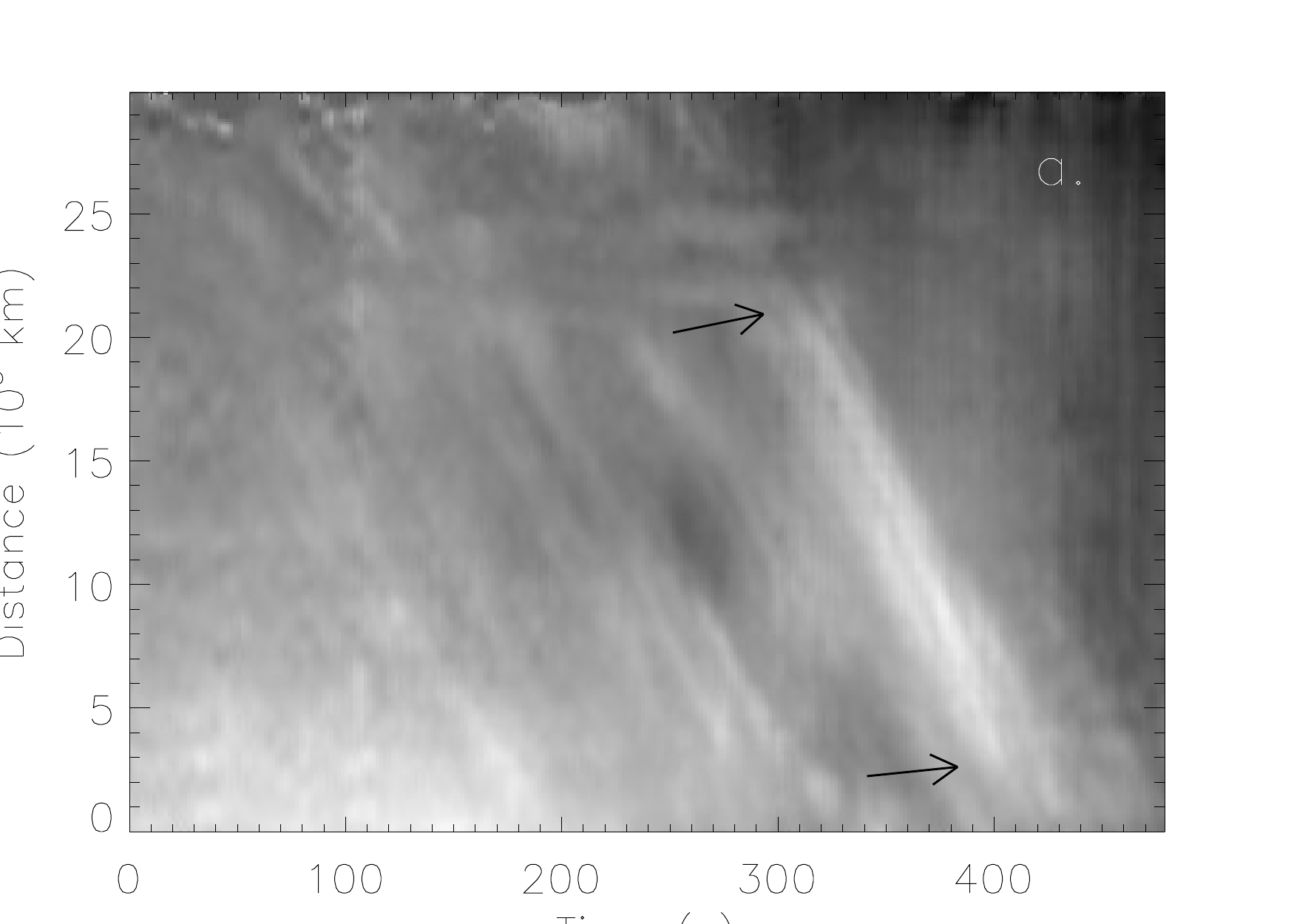}
\includegraphics[scale=0.50, angle=0]{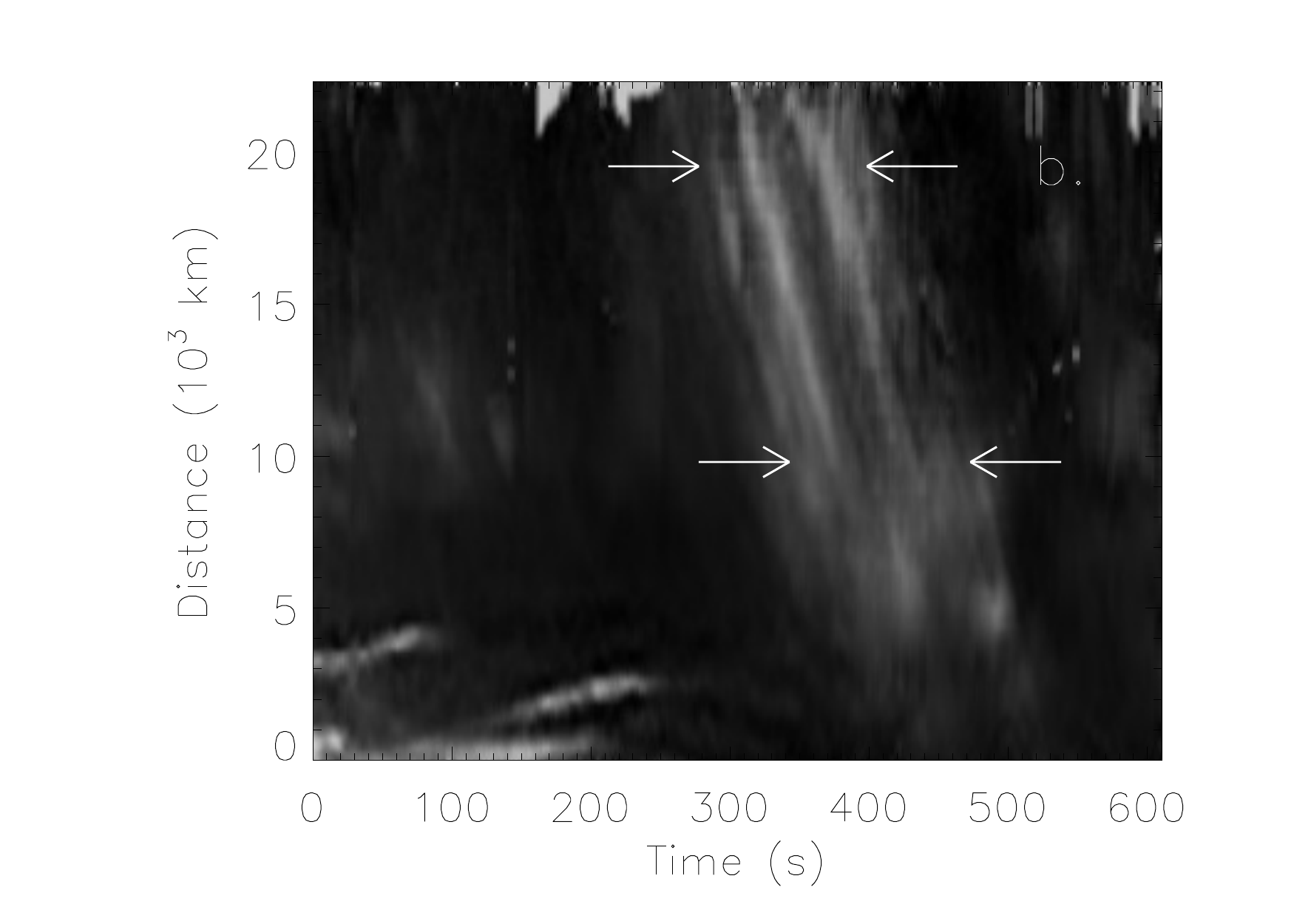}
\includegraphics[scale=0.55, angle=0]{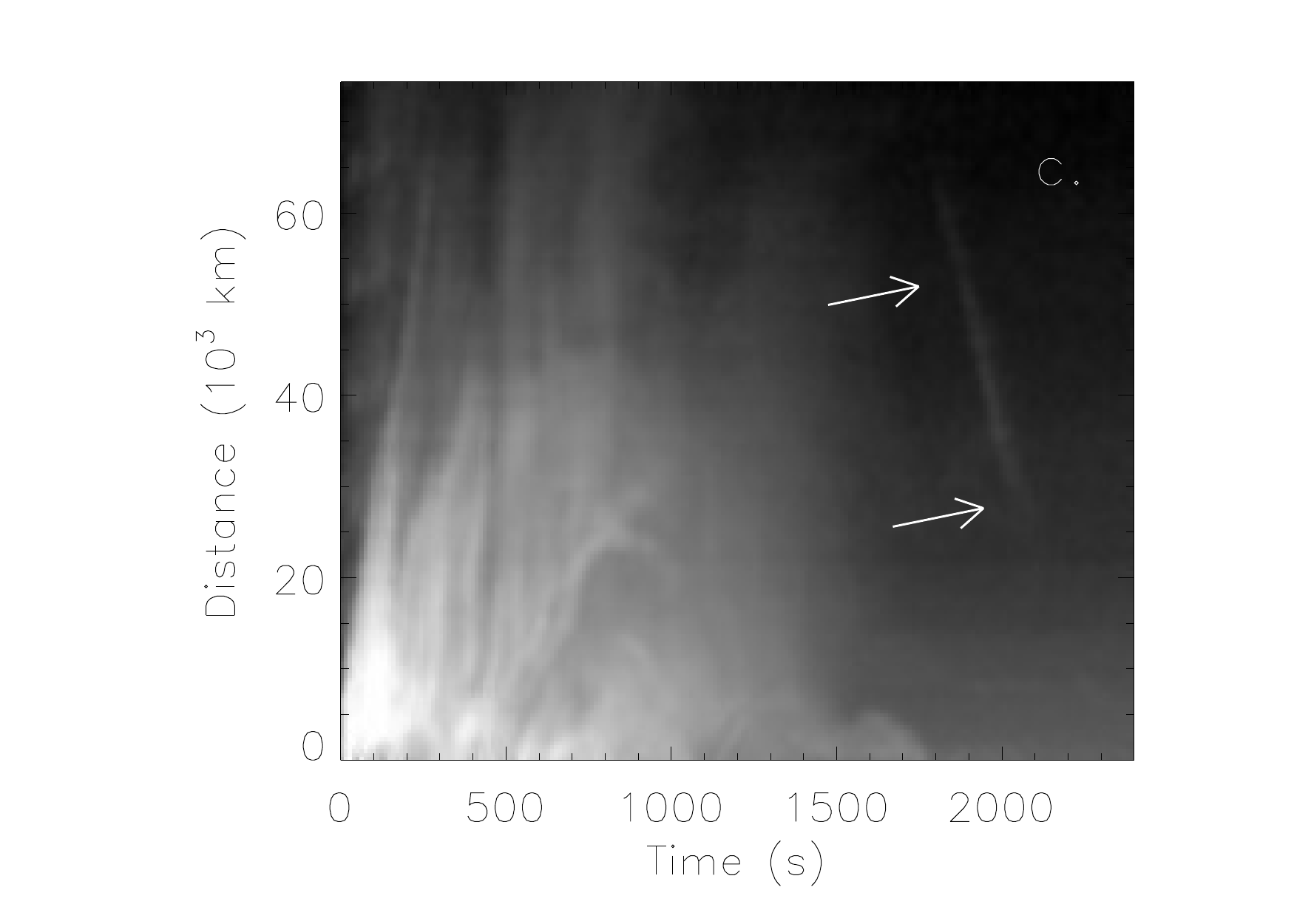} 
\caption{Space-time plots in H$\alpha$ and SDO 304\AA\  images from 11 Feb 2012 showing blobs falling back toward the solar surface. The top panels show the space-time plots for H$\alpha$ with: a. the left panel showing the time-distance plot for series 1  (UT 16:11:24 start time), and, b. the right panel shows the second series (start time UT 16:21:42). c. The lower panel shows the space-time plot of the SDO 304 \AA\ band starting at UT 16:00 with a logarithmic intensity scale. Arrows indicate the region from which velocities were derived (see text).
}
\label{xtcut}
\end{figure}
\clearpage

 \section{Results}
 Our H$\alpha$  observations detected many plasma condensations (``blobs") falling back toward the solar surface.  
We measured the condensation distances, velocities, and  sizes from the best sequence of H$\alpha$  images starting near 16:16:24 UT. 
 The largest plasma condensation, or  ``blob", entered the HARDcam field at $\sim$35,000 km from the solar limb at 16:16 UT. The average width  (direction perpendicular to its motion) and length (direction parallel to its motion) blob sizes were 5 and 30 pixels, corresponding to physical sizes of $\approx$500 and 3000 km, respectively.   This is larger than the typical sizes for coronal rain, a phenomena that will be further discussed in \S\ 4. 
Sample ``blob" sizes (taken from Figure~1) are shown in Table~1.
 
 We created space-time cuts of the H$\alpha$ images and found velocities of 196 $\pm$10 km/s (Figure~\ref{xtcut}), derived from the slope of the maximum length of the track in the X-T plot. These values are at the higher end for typical velocities found for CMEs and in coronal rain \citep{A10, AVR12a, ARdV12b, D14, I12, T06, T07, P13, X14}.
Velocities derived from the second set of H$\alpha$ observations were found to be 209$\pm$15 km/s and 218$\pm$20 km/s for the left and right space-time cuts in Figure 6b. 
The seeing becomes very poor near the end of the second set of  H$\alpha$ observations (near $\approx$16:28 UT), and this later interval of data and periods of poor seeing were not included in the velocity estimates.
\citet{K13} observed a failed CME with downwards velocities of 60$\pm$10 km/s and \citet{J03} 
found much higher typical velocities of $\approx$200 km/s.  
 
 We also observed several plasma condensations or blobs returning toward the solar surface in SDO AIA 304 \AA\ images (See Figure~\ref{sdo304}). The first blob during the H$\alpha$ sequence appeared at $\approx$16:15 UT at a distance of $\approx$40,000 km above the solar surface. Its width and length were 7$\arcsec \times 22\arcsec$, corresponding to a physical size of $\approx$5000 by 17,000 km$^2$.  Three other condensations were also observed by SDO during the second HARDCam sequence, and had sizes slightly larger than the first,  on the order of $\approx$7000 by 18,000 km$^{2}$. 
 In Table 1, we present a summary of the sizes of several of the condensation features (blobs) returning to the solar surface.
 
Velocities for the blobs were found to be $\approx$190 -- 200 km/s over the 10 minute sequence from 16:15 to 16:30 for the SDO AIA 304 \AA\ bandpass.  A velocity of $\approx$190$\pm$22 km/s was found for the first minute after the blob appeared at 16:17 UT and is in agreement with the velocities derived from the H$\alpha$  observations. The derived velocities are affected by the large cadence time of 12 seconds and poorer spatial resolution of the SDO. The SDO 1600~\AA\ band had a similar structure to the H$\alpha$ images and 
velocities of falling blobs ranged 180 -- 210 km/s, although the same blobs were not matched for both data sets. This may be the result of different temperatures and opacities observed in each band pass and  some information may be lost as a result of the 24 s cadence in the 1600 \AA\ band.

The start of the eruption, as seen in the SDO AIA filters, is marked in Figure 5, the time-distance diagrams, as a dashed line for Hinode/EIS observations.
Line widths observed with Hinode/EIS 
show an increase of Doppler shifts throughout the slit a few minutes later. 
Clear changes of Doppler shifts, from redshifts of +5 km/s to blue shifts of -5 km/s are observed all along the slit, for both the off-limb and the on-disk regions. Such relatively strong changes over a distance of over 500$\arcsec$ along the solar surface suggest a large scale reconfiguration of the magnetic field caused by the eruption. 
Particularly, in the slit range above 550$\arcsec$ we notice a dimming region in the intensity images, 
co-located to relatively strong variations of the Doppler shifts, from $+$10 km/s to $-$10 km/s over a time interval of 15 min. Within this interval, and also co-located, we observe strong variations in line width, with shifts from 40 to 70 km/s over short intervals of 5 min.
 Despite of the poor signal-to-noise, an increase in the Fe XII 186.74 line width is also observed in the same spatial and temporal locations. The time interval in which these variations are observed corresponds to the first part of the eruption, in which most of the material is seen going upwards. Correspondingly, the erupting material appears in emission in the cool AIA filters such as 304  \AA\ and 1600  \AA\, but appears in absorption in the 193  \AA\ filter, matching the dimming observed in the EIS Fe XII line. In the second part of the eruption, lasting $\approx$40 min, in which material is seen to fall back to the solar surface, the Fe XII 195 \AA\  line intensity is gradually increased and the Doppler velocity shows smaller variations of a few km/s are around the rest wavelength. Gradually over this 40 min interval the flows turn into constant redshifts of +5 km/s, the same values prior to the eruption. 
 The line width on the other hand decreases sharply from its maximum value in a time interval of 10 min and retain henceforth the same pre-eruption values of 40 km/s.

The sharp changes in line width are probably mainly caused by a non-thermal component, since no abrupt temperature increase is observed in Fe XII 195 \AA. Furthermore, the Doppler velocity also displays a significant change over the same time interval, supporting this scenario. A close inspection to the AIA filters, especially the 304 \AA\ filter, shows regions exhibiting both upward and downward flowing material over short intervals of time, and also transverse swaying. 
These dynamics can very well explain the increase in line width. The lack of prominence activation (as brightening in hotter pass bands) may be a common feature of such failed eruptions.

 \section{Discussion}
 
Failed CMEs, when the plasma does not escape from the Sun, promise to provide constraints on the physical parameters of the corona and upper chromosphere.
We have detected the phenomenon of plasma returning to the solar surface in the aftermath of a failed CME. These features were detected in high resolution H$\alpha$
observations of the solar limb, in addition to contemporaneous SDO/AIA data.
The SDO observations also observed the eruption, which we classify as a  failed prominence eruption, since prominence material being ejected from the Sun is not common in a surges, sprays, or jets.

\subsection{Free-fall Velocities}
\citet{ARdV12b},  using the CRisp Imaging SpectroPolarimeter (CRISP) in the H$\alpha$ band studied a large sample of blobs and found typical widths and lengths of 300 and 700 km, respectively. Our blobs  (condensation feature) are $\approx$2 and 5 times larger in width and length, respectively. 
This implies our blobs are not coronal rain, but more likely a failed or partly-failed CME.
This suggests important morphological differences between post-eruptive fallback prominence material and coronal rain in both the longitudinal and transverse direction to the flow. 
 Space-time cuts of the SDO AIA 304 \AA\ band found CME out-bound velocities of 200 to 300 km/s, which are consistent with CME velocities in the lower corona at distances less than 1$R_{\Sun}$ \citep{P13}. Our observed in-falling velocities range from 190 to 220 km/s for our H$\alpha$ observations and  $\approx$180 -- 210 km/s, for the SDO 304 and 1600 \AA\ bands. 
 
 The expected free fall velocity can be given by


\begin{equation}
     v_{ff} = \sqrt{\frac{2GM_{\sun}h_{max}}{R_{\sun}(R_{\sun} + h_{max})}} 
\end{equation}

\noindent where M$_{\sun}$ is the mass of the Sun, R$_{\sun}$ is the Sun's radius, and
h$_{max}$ is the maximum loop height \citep{M05}. 
From our sequence of HARDcam images we find h$_{max}$ to be about 350 pixels, or $\approx$35,000~km, and results in an expected free-fall velocity using the above equation of 135 km/s 
However, for the much large SDO field of view, we find h$_{max}$ to be $\approx$180,000~km, and this corresponds to a free-fall velocity of 280 km/s.  If the plasma observed in H$\alpha$ originated from these heights, then we would expect it to reach a free-fall velocity of $\approx$280 km/s.  We measure an inclination for blobs observed returning to the solar surface in our Sequence 1 to be falling at $\approx$25 degrees to the solar normal and $\approx$15--20 degrees in sequence 2.   We also note that any inclination of the guiding magnetic field to the normal to the solar surface would lower the effective gravitational acceleration. 
There is also the unknown inclination of the material to the observer's line-of-sight and this would imply we only observe a lower limit to the actual velocity. 


\begin{center}
\begin{figure}[t]
\includegraphics[scale=0.66, angle=0]{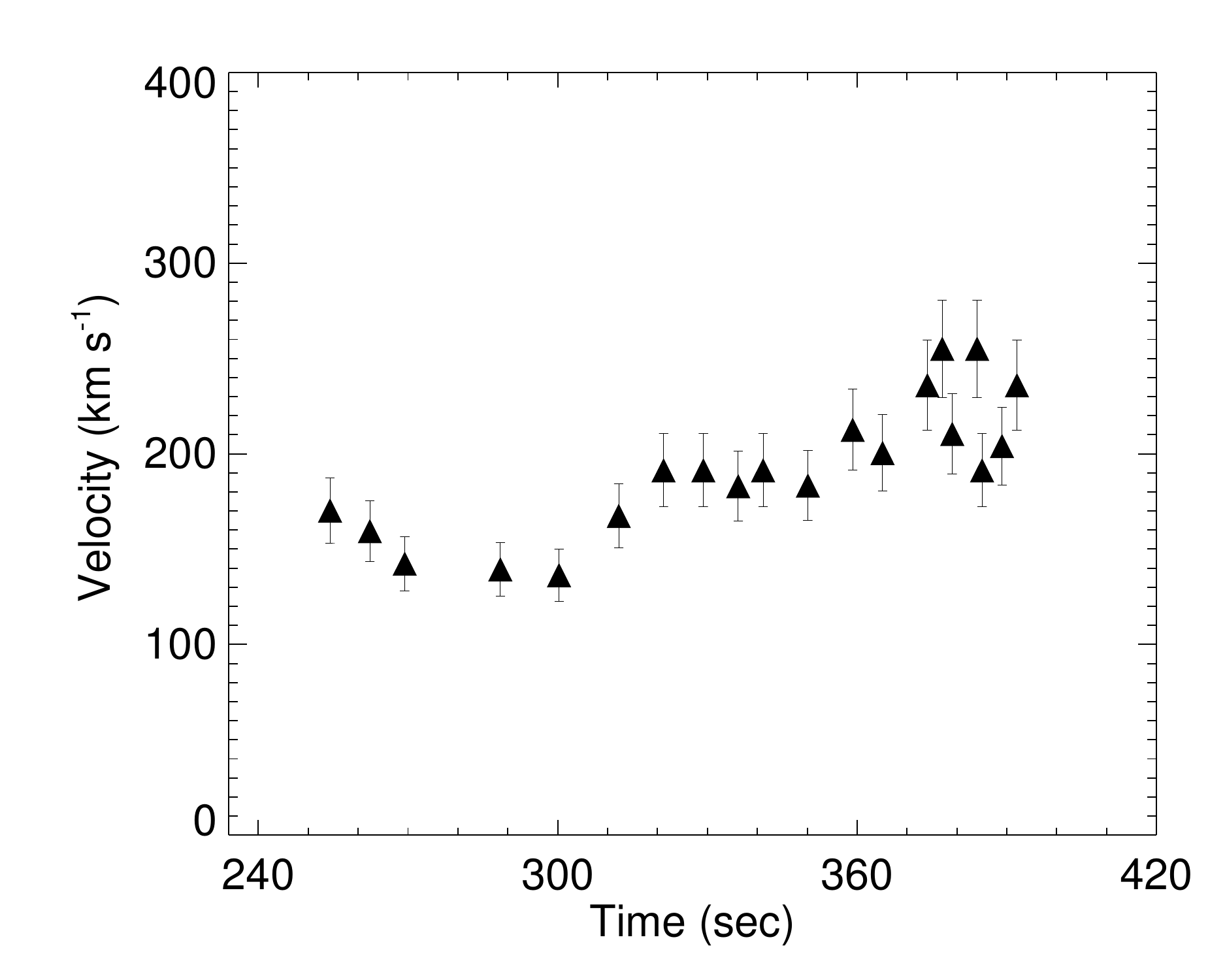}
\caption{H$\alpha$ velocity as a function of time for our first sequence of H$\alpha$ observations (Series 1) for the condensation H1, observed at 16:16:34 in the upper panel of  Figure 1 and in the space-time cuts in Figure 3.}
\label{vel}
\end{figure}
\end{center}

\begin{center}
\begin{figure}[t]
\includegraphics[scale=0.66, angle=0]{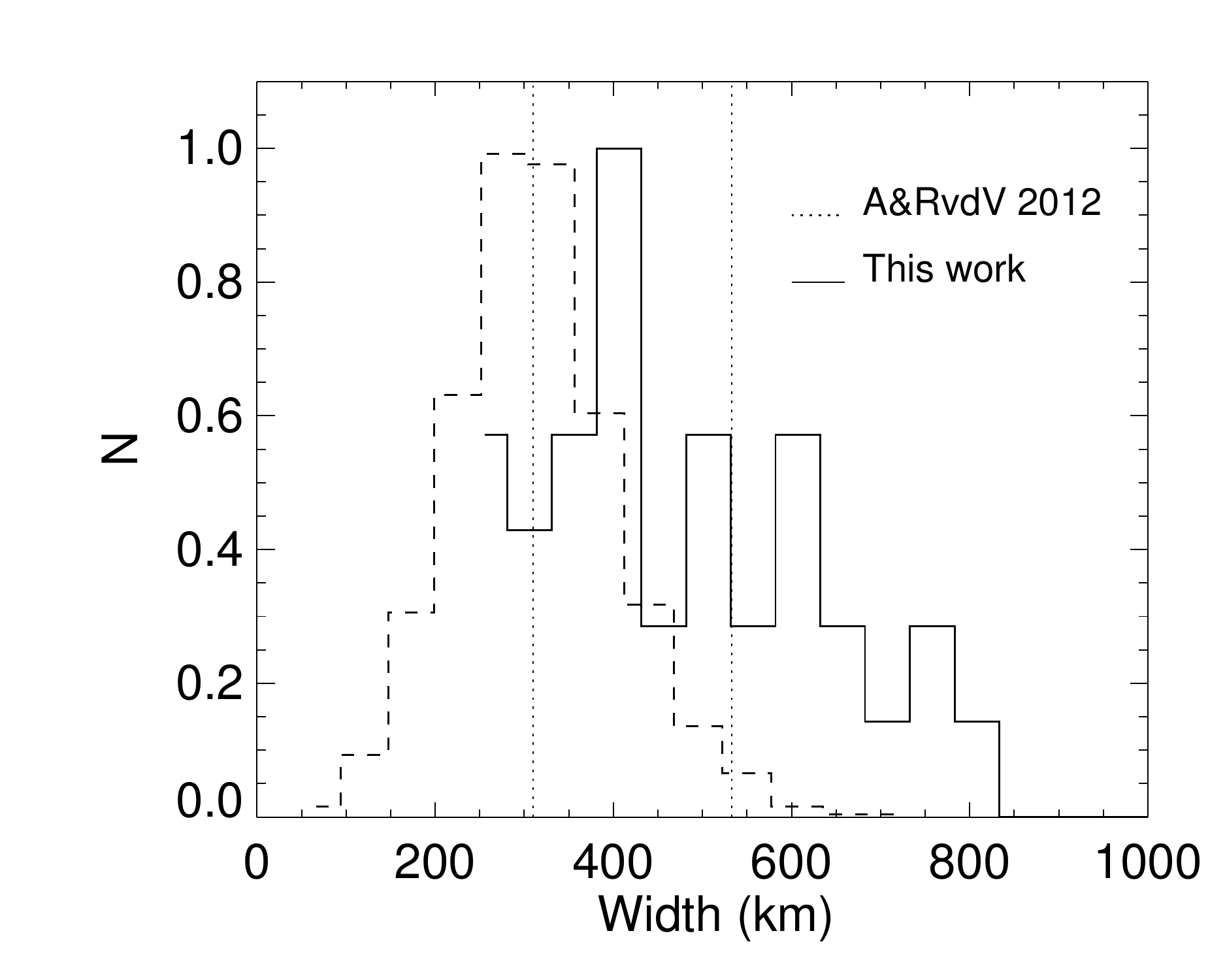}
\caption{Normalized histograms for the widths of the condensation observed in H$\alpha$. The histogram adapted from \citet{ARdV12b} is shown as the dashed line  (noted at A\&RvdV) and has an average width of 310 km (indicated by the horizontal dotted line). The distribution of the H$\alpha$ condensations measured in the work is shown as the solid line and has an average width of $\approx$530 km (indicated by the horizontal dotted line), owing to our larger resolution (see text).}
\label{hablobsize}
\end{figure}
\end{center}

Our first (Series 1) and second (Series 2) sequences of H$\alpha$ observations find free-fall velocities $\approx$20--30\% lower 
than those expected from gravity alone. Correcting for the small inclination angle between the returning material and the observer's line-of-sight decrease this disparity, but it still remains at 10-20\%. 
The observed velocities at values lower than the those expected from gravitational free-fall are in agreement to coronal rain observations \citep{S01,dG04,dG05,M05, A10, ARdV12b},
where average blob accelerations are 1/3 or less than solar gravity. Interpretations have been provided in terms of gas pressure variation from upward propagating slow modes \citep{A10} or pressure restructuring by slow modes produced by the blobs themselves. For the latter, slow mode shocks would travel down restoring the gas pressure and making the blobs fall at constant speeds \citep{O14}.

Ji et al. 2003 observed a maximum free-fall velocity of material returning to the solar surface after a failed filament eruption of nearly 300 km/s and 
found a maximum deceleration of this material of 10 times g$_{\sun}$ and interpret this force as magnetic tension.  The material we observed returning to the solar surface has a maximum velocity of $\approx$255 km/s.
and we find typical decelerations of the falling blobs of 2--3 g$_{\sun}$ with a maximum of $\approx$12 g$_{\sun}$ from the velocities presented in Figure 7. These velocities and decelerations were derived from measuring the slope along the main track in the X-T plot (Figure~\ref{xtcut}) at $\approx$7 second intervals. These decelerations are similar to those found by Ji et al., but we caution there is a large amount of scatter in the velocity measurements, there is the unknown inclination to the observer's line-of-sight, and the uncertainties may be large.

\subsection{Condensation Parameters}
Condensations (blobs)  returning to the solar surface are first observed at over $\approx$180,000~km above the solar surface in both the 304 \AA\ and 1600 \AA\ bandpasses. 
The 1600 \AA\ continuum bandpass is dominated by emission from C IV with a temperature of about 10$^5$ K.
The 304 \AA\ channel is sensitive to He II emission at  $\approx5\times10^4$ K, however the off-limb emission can be dominated by  Si~XI emission with a temperature of 1.6 $\times$10$^6$ K \citep{O10}. 
This material is detected at $\approx$35,000 km above the solar surface in the H$\alpha$ band (temperature $\approx$ 10$^4$ K). Although similar structures are observed in the 1600 \AA\ and H$\alpha$ bandpasses (See Figure 2 and Movie 1), we were unable to match blobs in the different bandpasses. This may be a result of both the very different temperatures and opacities in the different bands and the spatial and temporal sampling.

Our H$\alpha$ observations have only detected $\approx$50 blobs returning to the solar surface, that are over 2$\sigma$ significance above the median intensity (the parameters for a sample of these are presented in Table 1).  We find  there is not so much clumpy structure in our observations.
\citet{ARdV12b}  detected nearly 4000 blobs returning to the solar surface in the CRISP observations of both on-disk and off-disk coronal rain, and found a mean width for their distribution for on-disk coronal rain of 310 km. We have computed a similar distribution for our H$\alpha$ observations and find a average width of about 530 km. We compare these two H$\alpha$ size distributions in Figure~\ref{hablobsize}.  Our larger average sizes are consistent with our factor of $\approx$2  larger resolution. We also note, as did \citet{ARdV12b}, that there may be many more unresolved structures, and the slight rise to smaller ``blob" widths in our distribution supports this.  The structure observed in coronal rain may form as the result of thermal instabilities and blob sizes may provide further constrains on the detailed physics of eruptions and coronal rain \citep{OT11, O14}.
\citet{F13} simulated 4000 ``blobs" returning to the solar surface in an 80 minute period and found an average width of 400 km, and note only about 25\% of these blobs would be detectable with the current instrumental resolution. Such a ``blob'' distribution is consistent with our H$\alpha$ observations.

\subsection{Failed CME Kinetic Energy}

The main outburst observed in the SDO bands starting at $\approx$15:55 UT and it was also accompanied by a strong increase of Doppler shifts (20 km/s change from the start of the eruption) and a strong increase in line width (from 40 to 70 km/s) as observed by the {\it Hinode} EIS.  We approximate the extent of the area of the CME at 16:06UT as a triangular shape of 6000 x 25300 km$^2$. Assuming its thickness is equal to its width, we find a volume of 2.4 $\times$ 10$^{11}$ km$^{3}$ and this gives a mass of  $\approx4.0\times10^{11}$ -- $10^{13}$ g for an estimated range of densities of 10$^9$ to 10$^{11}$ cm$^{-3}$ spanning values for typical densities in prominences to flares  \citep{H85,O10}. This mass gives a maximum kinetic energy in the main eruption of $\approx$2$\times10^{28}$ ergs for a velocity of 300 km/s, much smaller than a typical large CME  \citep{F00}.   These values are two orders of magnitudes lower than the typical CME masses of $10^{15}$ -- $10^{16}$ g \citep{G07}.  This low kinetic energy may explain why this is a partial CME. We now derive masses of our of typical blobs using a volume from the average blob size 500$\times$3000 km$^{2}$ and assuming a thickness equal to our measured 500 km width, and an average density,  gives masses of $\approx1.3\times10^{10}$ g. 
Such a blob would have a kinetic energy of  $\approx5.0\times10^{24}$ ergs 
for this average mass, and this is also several orders of magnitude smaller than those found from typical CMEs \citep{F00}.  Additionally, if we sum up the masses of the $\approx$50 individual blobs (including those presented in Table~1), 
$\rho\sum\limits_{i=1}^n m_i$,
we find a total mass of $\approx5.0\times10^{11}$~g for an average density of 10$^{10}$ cm$^{-3}$. This implies only about 18\% of the material returns to the solar surface as compared to our estimated mass in the main outburst. If we have detected only about half of the blobs returning to the solar surface, then the total mass is still less the 36\% of the initial CME and the majority of mass escapes from the Sun. Our SDO, H$\alpha$ and EIS observations of this failed CME support the outburst observed is consistent with a partial eruption as described in  \citet{G07}, and only a fraction of the ejected mass returns to the solar surface in the form of blobs.

\section{Concluding Remarks}
\label{conc}
We have used a multi-instrument, multi-wavelength approach to obtain high spatial and temporal resolution observations of the upper solar chromosphere and have detected condensations (``blobs'') returning to the solar surface at velocities of $\approx$200 km/s after a failed prominence eruption. 
Velocities in the upper corona from SDO AIA images were in general agreement, but blobs could not be matched to the ones observed in H$\alpha$, probably as the result of very different temperature and opacity distributions, and also spatial and temporal sampling. Derived  average ``blob"  sizes in H$\alpha$ are $\approx$ 500 x 3000 km$^{2}$ in the directions perpendicular and parallel to the direction of travel, respectively and are much larger than sizes found for typical coronal rain. A comparison of our ``blob" widths to those found from coronal rain,  indicate there are additional smaller, unresolved ``blobs" in agreement with recent numerical simulations \citep{F13}. 
We interpret our H$\alpha$ and EUV observations as a partial CME. We derived a kinetic energy $\approx$2 orders of magnitude lower for the main eruption (at UT 15:55) than a typical CME, which may explain its partial nature.
Our observed velocities and decelerations of the blobs  in both H$\alpha$ and SDO bands are less than those expected for gravitational free-fall and
imply additional magnetic or gas pressure impeding the flow.  However, we realize additional high spatial and temporal resolution observations of the solar limb are needed to quantify both failed CMEs and other phenomenon which will allow further constraints on the coronal magnetic field and in the larger problem of coronal heating.

\acknowledgements
We thank D. Gilliam and all of the DST staff for their excellent 
support for this project.  We thank an anonymous referee for suggested improvements to this manuscript. DC would like to dedicate this paper to his late mother, RoseMarie Sciortino Christian.

\begin{landscape}
\label{obs}
\tabletypesize{\small}
\begin{deluxetable}{lccccccccccr}
\tablewidth{0pt}
\tablenum{1}
\tablecaption{Sample Condensation Features}
\tablehead{
   \colhead{Feature}
  &\colhead{Time}
  &\colhead{Size}
  &  \colhead{}
  & \colhead{Feature}
  &\colhead{Time}
  &\colhead{Size}
  &  \colhead{}
  & \colhead{Feature}
  &\colhead{Time}
  &\colhead{Size}
  &\colhead{Comments\tablenotemark{a}}
  \\
  \colhead{}
  &\colhead{(UT)}
  &\colhead{(km$^2$)}
      &\colhead{}
       &\colhead{}
  &\colhead{(UT)}
  &\colhead{(km$^2$)}
   &\colhead{}
    &\colhead{}
  &\colhead{(UT)}
  &\colhead{(km$^2$)}
   &\colhead{}}
   \startdata 
 \multicolumn{3}{c}{H$\alpha$} & &  \multicolumn{3}{c}{AIA 1600 \AA} & &  \multicolumn{3}{c}{AIA 304 \AA} \\
     \cline{1-3} 
      \cline{5-7} 
        \cline{9-11}
                    H1 & 16:16:34  & 700 x 1300  & 
                   & A1 &  16:16:41 &  2350 x 9960  &
                   & B1 &  16:16:44   & 1090 x   9000      \\   
                     H2 & 16:16:51  & 730 x 2900  &  
                      & A2a &  16:22:41 &  1830 x 8480   &
                    & B2a & 16:22:44 &  2910 x 9050  \\
                              &                &                &     
                    & A2b &  16:22:41 &   4180 x 2700   & 
                   & B2b & 16:22:44 & 2610 x 1830  & horiz.  \\ 
                            &                &                &     
                   & A2c &  16:22:41 &  1090 x 3050   &
                  & B2c & 16:22:44 &  1600 x 3600 \\
                     H3 & 16:17:55 & 770 x 4340 &  
                   & A3a  &  16:23:53 &     2910 x  9270   &   
                    & B3a & 16:23:56 &  1670 x 10530 \\
                        &                &                 &    
               & A3b  &  16:23:53 &      1440 x 1780  &
               & B3b & 16:23:56 &  1610 x 1520  \\
                         &                &           &          
                & A3c  &  16:23:53 &       1300 x 2180  &
               & B3c & 16:23:56 &    1440 x 2130 \\
                H4  & 16:18:41  &  500 x 2340 &
               & A4a   & 16:29:53 &      1650 x 13880      & 
               & B4a  &  16:29:44 &   1830 x 6310   \\
                     &                &            &          
                                  & A4b   & 16:29:53 &      4390 x 2570       &
             & B4b  &  16:29:44 &   3610 x 1520  & horiz.  \\

                      &                &                 &    
             & A4c   & 16:29:53 &       2570 x 1220    &
             & B4c  &  16:29:44 &    1610 x 870   & horiz. \\
               H5a & 16:23:53 &  410 x 2620    \\
               H5b & 16:23:53 &  410 x 2790   \\
                H6a & 16:25:47 & 1260 x 4900 & 
                      &               &                &  &
                    &   &               &                & 2 blobs \\  
                H6b & 16:25:47 &  470  x 2300             \\
                          & H$\alpha$   \\
              \cline{1-3}\\
              H7a & 16:26:10 &  1170 x 6900 & 
                      &               &                &  &
                    &   &               &                & 2 blobs \\ 
              H7b   &  16:26:10 &  410 x 1740              \\
                  H8a & 16:26:42 &  590 x 3450 \\
             H8b & 16:26:42 &  340 x 5330  \\
             H9a & 16:27:50 &  830 x 4600 \\
            H9b & 16:27:50 &  340 x 5600  \\ \hline
\tablenotetext{a}{Comments: horiz.  = a horizontal feature in SDO; 2 blobs = evidence for two unresolved blobs. }
\enddata
\end{deluxetable}
\end{landscape}



\end{document}